\documentclass[journal]{IEEEtran}
\usepackage{amsmath}
\usepackage{amssymb}
\usepackage{amsfonts}
\usepackage{graphicx}
\usepackage{epsfig}
\usepackage{epstopdf}
\usepackage{subfigure}
\usepackage{psfrag}
\usepackage{cite}
\usepackage{url}
\usepackage{color}
\usepackage{bigstrut}
\usepackage{multirow}
\usepackage{array}
\usepackage{authblk}
\usepackage[dvipsnames]{xcolor}
\usepackage{accents}
\newcolumntype{N}{>{\centering\arraybackslash}p{5em}}
\newcolumntype{D}{>{\centering\arraybackslash}p{7em}}
\newcolumntype{S}{>{\centering\arraybackslash}p{5em}}

\title{Waveform Optimization for Radio-Frequency  Wireless Power Transfer}
\author{\vspace{-1mm}Mohammad R. Vedady Moghadam, Yong Zeng, and Rui Zhang} 
\affil[]{\vspace{-3mm}ECE Department,  National University of Singapore. E-mail:\{elemrvm, elezeng,  elezhang\}@nus.edu.sg \vspace{-6mm}}
\begin{document}
\maketitle \thispagestyle{empty}
\pagestyle{empty}
\begin{abstract}
In this paper, we study the waveform design problem for a single-input single-output (SISO) radio-frequency (RF) wireless power transfer (WPT) system in frequency-selective channels. First, based on the actual non-linear current-voltage model of the diode at the energy receiver, we derive a semi-closed-form expression for the deliverable DC voltage in terms of the incident RF signal and hence obtain the average harvested power.  
Next, by adopting a multisine waveform structure for the transmit  signal of the energy transmitter, we jointly design the multisine signal  amplitudes and phases over all frequency tones  according to  the channel state information (CSI) to maximize the deliverable DC voltage or harvested power. 
Although our formulated problem is non-convex and difficult to solve, we propose two suboptimal   solutions to it, based on the frequency-domain maximal ratio transmission (MRT) principle and  the  sequential convex optimization (SCP) technique, respectively.
Using various simulations, the performance gain of our solutions over the existing  waveform designs is shown.
\end{abstract}
\vspace{-1mm}
\newtheorem{definition}{\underline{Definition}}[section]
\newtheorem{fact}{Fact}
\newtheorem{assumption}{Assumption}
\newtheorem{theorem}{\underline{Theorem}}[section]
\newtheorem{lemma}{\underline{Lemma}}[section]
\newtheorem{corollary}{\underline{Corollary}}[section]
\newtheorem{proposition}{\underline{Proposition}}[section]
\newtheorem{example}{\underline{Example}}[section]
\newtheorem{remark}{\underline{Remark}}[section]
\newtheorem{algorithm}{\underline{Algorithm}}[section]
\newcommand{\mv}[1]{\mbox{\boldmath{$ #1 $}}}
\vspace{-4mm}
\section{Introduction}\label{Section:Introduction}
Radio frequency (RF) wireless power transfer (WPT) is a promising technology to provide convenient and sustainable power supply to low-power devices \cite{888}. 
Different from the near-field WPT techniques, e.g.,  inductive coupling \cite{671} and magnetic resonant coupling \cite{501,840}, RF WPT utilizes the far-field electromagnetic (EM) radiation for remote power delivery, which has many promising advantages such as longer power transmission range, smaller receiver/transmitter form factors, applicable even in non-line-of-sight (NLoS) environment, easier implementation of power multicasting to a large number of devices simultaneously, etc.

Early work on RF WPT has been historically targeted for long-distance and high-power transmissions, as mainly driven by the two appealing applications of wireless-powered aircraft and solar power satellite (SPS). During the past decade, the interest in WPT has been mostly shifted to enable relatively low-power delivery over moderate distances  due to the increasing need for remotely charging various devices such RFID tags, Internet of Things (IoT) devices, wireless sensors, etc. 
Remarkably, tremendous research efforts have been recently devoted to the study of WPT for applications in wireless communications, by exploiting the dual usage of RF signals for carrying energy and information. 
There are mainly two lines of research along this direction, namely {\it simultaneous wireless information and power transfer} (SWIPT) \cite{478}, where information and power are transmitted concurrently using the same RF signal in the same direction, and {\it wireless powered communication} (WPC) \cite{515}, where the energy for wireless communication at the devices is obtained via WPT. 
More recently, there have been increasing interests in applying advanced communications and signal processing techniques for designing efficient WPT systems, such as energy beamforming  via efficient channel estimation \cite{491}, \cite{528}, multi-user charging  scheduling \cite{813,835}, massive MIMO \cite{889,843} and millimeter wave technologies \cite{833}, etc.

However, all the aforementioned work assumed the linear energy harvesting (EH) model, i.e., the RF-to-direct current (DC) power conversion efficiency of the \textit{rectenna} (i.e., a receive antenna combined with a rectifier that typically consists of a diode and a low pass filter (LPF)) at the energy receiver is assumed to be constant regardless of its incident RF signal power and waveform.  
Though providing a reasonable approximation for extremely low incident power at the rectenna, the linear model is inaccurate in most practical scenarios. 
On one hand, the RF-to-DC power conversion efficiency typically increases with the input power, but with diminishing returns and eventually saturates due to the diode reverse breakdown \cite{Boshkovska:2015}. 
On the other hand, even with the same input RF power, the power conversion efficiency in fact critically depends on the actual RF waveform \cite{Trotter:2009,544,Clerckx:2016b}. Experimental results have shown that signals with high peak-to-average power ratio (PAPR), such as the OFDM signal or chaotic waveforms, tend to result in more efficient RF to DC power conversion \cite{544}. 
The practical rectenna non-linearity thus has a great impact on the design of end-to-end WPT systems, which, however, was not rigorously investigated before until the landmark work  \cite{Clerckx:2016b}. 
In \cite{Clerckx:2016b}, the authors studied the multisine  waveform design problem for RF WPT systems, where the rectenna non-linearity is approximately characterized  via the second and higher order terms in the truncated Taylor expansion of the diode output current. 
Based on this model, a sequential convex programming (SCP) algorithm is proposed to approximately design the amplitudes of different frequency tones in an iterative manner, where at each iteration a geometric programming (GP) problem needs to be solved.

In this paper, we study the waveform optimization problem for a single-input single-output (SISO) WPT system in frequency-selective channels  to fully exploit the nonlinear EH model in the multisine waveform design to maximize the end-to-end efficiency.  
Different from the prior work  \cite{Clerckx:2016b}, we first develop a generic EH model based on circuit analysis that accurately captures the rectenna nonlinearity without relying on Taylor approximation as adopted in \cite{Clerckx:2016b}. 
By assuming that the capacitance of the LPF of the rectenna is sufficiently large (similar to \cite{Clerckx:2016b}), our new model shows that maximizing the DC output power is equivalent to maximizing the time average of an exponential function in terms of the received  signal waveform. 
Based on this new model, we then formulate a new multisine waveform optimization problem subject to the transmit sum-power constraint. 
Two approximate solutions are then proposed for our formulated problem, which is non-convex in general and thus difficult to solve optimally. 
The first solution, which is given in closed-form, essentially corresponds to a maximal ratio transmission (MRT) over the frequency tones. 
This is in a sharp contrast to the conventional linear EH model, for which all transmit power should be allocated to a single frequency tone with the strongest channel gain \cite{888}. 
In the second proposed solution, we employ an SCP based algorithm to iteratively search the optimal amplitudes of the multisine signal, where at each iteration the problem is approximated by a convex quadratically constrained linear programming (QCLP), for which the optimal solution is derived in closed-form and thus can be efficiently computed. 
Hence, compared to the existing SCP-GP  algorithm in \cite{Clerckx:2016b}, our proposed SCP-QCLP algorithm significantly reduces the  computational complexity, which is confirmed by our simulations. 
Moreover, the proposed algorithm guarantees to converge  to (at least) a locally optimal solution satisfying the KKT (Karush-Kuhn-Tucker) conditions of our formulated waveform optimization problem.  

The rest of this paper is organized as follows. 
Section II introduces the system model.  
Section III presents  the rectenna circuit analysis. 
Section IV  formulates the multisine waveform optimization problem, and presents two approximate solutions to it. 
Section V shows the performance of our proposed designs.  
Finally, we conclude the paper in Section VI.
\vspace{-2mm}
\section{System Model}\label{Section:system_model}
\begin{figure}
	\centering
	\includegraphics[width=0.74\linewidth]{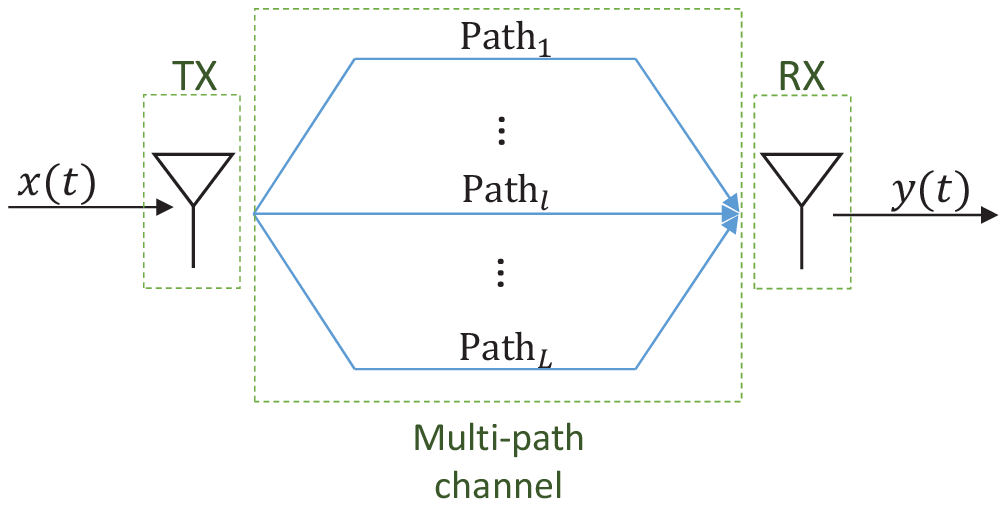} \vspace{-3mm}
	\caption{System model.}  
	\label{fig:System_Model} \vspace{-4mm}
\end{figure}

As shown in Fig. \ref{fig:System_Model},  we consider a point-to-point  WPT system  where an energy transmitter is intended to deliver energy wirelessly to an energy receiver, which is known as {\it rectenna}.  
To reveal the most essential design insights, we assume that both the transmitter and receiver are equipped with a single antenna, while our design method can be similarly applied  for the multi-antenna case. 
We assume that the available frequency band for WPT is continuous and given as $[f_{\min},~ f_{\max}]$,  with $f_{\min}>0$ and $f_{\max}>f_{\min}$  in Hz.  Accordingly, we  define   $B=f_{\max}-f_{\min}$ and $f_c=(f_{\max}+f_{\min})/2$ as  the total bandwidth and the central frequency, respectively. 
Similar to the multisine waveform structure in  \cite{Clerckx:2016b}, we assume that $N\ge 1$  sinewaves are being used for WPT. We set their  frequency tones as $f_n=f_0+ (n-1) \Delta_f$, $n=1,\ldots,N$, where $f_0\ge f_{\min}$ and $\Delta_f>0$ are designed such that $f_0/ \Delta_f$ is an integer and $f_0+(N-1)\Delta_f \le f_{\max}$. In particular, we  set $\Delta_f= B/N$ and $f_0=\lceil f_{\min}/\Delta_f\rceil \Delta_f$, with $\lceil a \rceil$ denoting the smallest integer greater than or equal to $a$.  
Hence, the transmit signal over time $t$  is expressed as  $x(t)=\Re\{\sum_{n=1}^{N} \sqrt{2} \tilde{s}_{n} \exp(jw_nt) \}$, with $w_n=2\pi f_n$ and $\tilde{s}_{n}=s_{n}\exp(j\phi_{n})$, where $s_{n}\ge 0$ and $0 \le \phi_{n}< 2 \pi$ denote the amplitude and phase of the $n$-th sinewave at frequency $f_n$, respectively.    
In the rest of this paper, we treat $s_{n}$'s and $\phi_{n}$'s as \textit{design variables}.
It can be verified that $x(t)$ is periodic, with the period $T=1/\Delta_f$.  
Moreover, the transmitter  is subject to a maximum power constraint, denoted by $P_T>0$, i.e., \vspace{-3.5mm}
\begin{align} \label{eq:Sum_Power_constraint}
\dfrac{1}{T}\int_{T} x(t)^2 dt=\sum_{n=1}^{N}  s_{n}^2\leq P_T.
\end{align}

We consider that the transmitted signal propagates  through a multipath channel, with $L\ge 1$ paths,  where the delay, amplitude, and phase  for each path $l$ are denoted by $\tau_l >0$, $\alpha_l >0$, $0 \le \xi_l < 2\pi$, respectively.
The signal received at the rectenna after multipath propagation is thus given by  \vspace{-2.5mm}
\begin{align} \label{eq:y(t)}
y(t)&= \Re \left\{\sum_{n=1}^{N} \sum_{l=1}^{L}\sqrt{2} s_{n}\alpha_l \exp(j(w_n(t-\tau_l)+\xi_l+\phi_{n}))\right\}\nonumber\\
&=\sum_{n=1}^{N} \sqrt{2} s_{n} h_{n} \cos(w_n t+\psi_{n}+\phi_{n}),
\end{align}
where $h_{n}$ and $\psi_{n}$ are the amplitude and phase of the channel frequency response at  $f_n$, such that $h_{n} \exp(j \psi_n)=\sum_{l=1}^{L}\alpha_l \exp(j(-w_n\tau_l+\xi_l))$. In this paper, we assume the CSI, i.e.,  $h_{n}$'s and $\psi_{n}$'s,  is known to the transmitter.  
\vspace{-1mm}
\section{Circuit Analysis of  Rectenna}\label{Section:Analytical_Model_Rectenna}
In this section, we present a simple and tractable nonlinear  model of the rectenna circuit, and derive its output DC voltage as a function of the received signal by the rectenna. 
\vspace{-3mm}
\subsection{Rectenna Equivalent Circuit}\label{Sect:Rectenna_eq_circuit}
\begin{figure}
	\centering
	\includegraphics[width=0.93\linewidth]{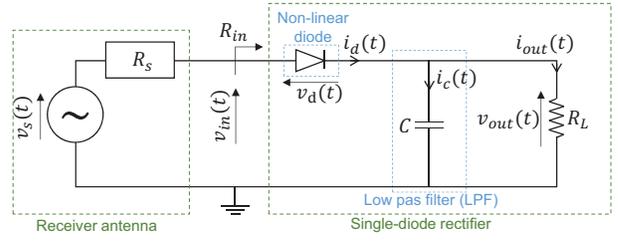} \vspace{-2mm}
	\caption{Rectenna circuit model.}
	\label{fig:antennafull} \vspace{-4mm}
\end{figure}

As shown in Fig. \ref{fig:antennafull}, a typical rectenna consists of two main components, namely an antenna collecting EM waves (i.e., RF signals) from the air,  and a single-diode  rectifier converting the collected signal to DC for direct use or charging a battery. 
The antenna is  commonly modelled as a voltage source $v_{s}(t)$ in series with a  resistance $R_{s}>0$, where $v_{s}(t)=2  \sqrt{R_{s}} y(t)$ \cite{Clerckx:2016b}, with $y(t)$ denoting the received signal  as given in (\ref{eq:y(t)}).  
On the other hand, the rectifier consists of a single diode, which is   non-linear,  followed by a LPF connected to an electric load, with the load resistance  denoted by  $R_{L}>0$.
As shown in Fig. \ref{fig:antennafull}, we denote $R_{in}>0$ as the equivalent input resistance of the rectifier. 
With the perfect impedance matching, i.e., $R_{s}=R_{in}$, the input voltage of the rectenna, denoted by $v_{in} (t)$, is obtained as $v_{in}(t)=v_{s}(t)/ 2= \sqrt{R_{s}} y(t)$. 
Since $y(t)$ is periodic, it follows that  $v_{in} (t)$ is also periodic with the same period $T$.

Let $i_d(t)$ and $v_d(t)$ denote the current passing through the diode and the voltage drop across it, respectively. 
By assuming that the diode is ideal, we have  
\begin{align} \label{eq:Diode_iv}
i_d(t)=I_0\left(\exp(\dfrac{v_d(t)}{\eta V_0})-1\right),
\end{align}
where $I_0>0$ is the reverse bias saturation current of the diode, $V_0>0$ is its  thermal voltage, and $\eta>0$ is the  ideality factor. 
\vspace{-3mm}
\subsection{Performance Analysis}\label{Section:Performance characterization} 
By applying Kirchhoff's circuit laws to the electric circuit of the rectifier shown in Fig. \ref{fig:antennafull}, we obtain \vspace{-1mm}
\begin{align} 
&i_d(t)=i_c(t)+ i_{out}(t),\label{eq:Krichhof1}\\ 
&i_c(t)=C \dfrac{d v_{out}(t)}{dt},\label{eq:Krichhof2}\allowdisplaybreaks \\
&v_{d}(t)=v_{in}(t)-v_{out}(t),\label{eq:Krichhof3}\\
&v_{out}(t)= R_L i_{out}(t), \label{eq:Krichhof4}
\end{align}
where $v_{out}(t)$ and $i_{out}(t)$ are the  output voltage and current of the rectifier, respectively, while  
$C>0$ and $i_c(t)$ are the capacitance of the LPF and the current passing through it, respectively.   
After some manipulation based on  (\ref{eq:Diode_iv})--(\ref{eq:Krichhof4}), the relation between the input and output voltages of the rectifier is obtained as
\begin{align} \label{eq:vout(t)-vin(t)}
\hspace{-2mm} I_0\left(\exp(\dfrac{v_{in}(t)-v_{out}(t)}{\eta V_0})-1\right)=C\dfrac{d v_{out}(t)}{dt} + \dfrac{v_{out}(t)}{R_L}.\hspace{-2mm}
\end{align}

Let  $v_{out}(t)=\overline{v}_{out}+ \tilde{v}_{out}(t)$, with  $\overline{v}_{out}$ and $\tilde{v}_{out}(t)$ denoting the DC and AC (alternating current)  components of the output voltage, respectively.  
Since the input voltage of the rectifier   $v_{in} (t)$ is periodic,  it follows from  (\ref{eq:vout(t)-vin(t)}) that $\tilde{v}_{out}(t)$ is also periodic with the same period $T$, i.e., $\tilde{v}_{out}(t)=\tilde{v}_{out}(t+kT)$ for any integer $k$. Moreover,  the average value of $\tilde{v}_{out}(t)$  is zero, i.e., $\tfrac{1}{T}  \int_{T} \tilde{v}_{out}(t) dt=0$. 
Next, by averaging both sides of (\ref{eq:vout(t)-vin(t)}) over the period $T$, we obtain \vspace{-1mm}
\begin{align} \label{eq:vout(t)-vin(t)-int}
\hspace{-2mm}&I_0 \left(\dfrac{1}{T} \exp(-\dfrac{\overline{v}_{out}}{\eta V_0}) \int_{T}  \exp(\dfrac{v_{in}(t)-\tilde{v}_{out}(t)}{\eta V_0})dt  -1\right) \nonumber \\
&= \dfrac{C}{T} \int_{T} \dfrac{d \tilde{v}_{out}(t)}{dt} dt + \dfrac{1}{T R_L} \int_{T}  \big(\overline{v}_{out}+\tilde{v}_{out}(t)\big) dt \nonumber \\
&= C\big(\tilde{v}_{out}(T) - \tilde{v}_{out}(0)\big)+ \dfrac{1}{T R_L} \big(\ T \overline{v}_{out} +0\big)=\dfrac{\overline{v}_{out}}{R_L}.
\end{align}
By assuming that the capacitance $C$ of the LPF is sufficiently large, the output AC voltage of the rectifier is small, i.e.,   $\tilde{v}_{out}(t)\approx 0$ \cite{Clerckx:2016b}.   
Hence,  (\ref{eq:vout(t)-vin(t)-int}) can be simplified as   \vspace{-1mm}  
\begin{align} \label{eq:DC_vout}  
\exp(\dfrac{\overline{v}_{out}}{\eta V_0})\left(1+\dfrac{\overline{v}_{out}}{R_L I_0}\right)&=\dfrac{1}{T} \int_{T}  \exp(\dfrac{v_{in}(t)}{\eta V_0}) dt \nonumber \\
&= \dfrac{1}{T} \int_{T} \exp(\dfrac{\sqrt{R_s}y(t)}{\eta V_0}) dt.
\end{align}
The DC power delivered to the load is thus  given by \vspace{-2mm}
\begin{align} \label{eq:pdc}
\overline{p}_{out}=\dfrac{\overline{v}_{out}^2}{R_L}. 
\end{align}

From (\ref{eq:DC_vout}), it is observed that its left hand side (LHS) is strictly increasing over $\overline{v}_{out}$. Thus,  maximizing the output DC voltage/power is equivalent to maximizing the right hand side (RHS) of (\ref{eq:DC_vout}) by optimizing the received signal  $y(t)$. 
Since $y(t)$ is a function of the transmitted signal $x(t)$, we can alternatively design $x(t)$ to achieve this goal. With the optimized $x(t)$ and the resulted $y(t)$, we can evaluate the integration on the RHS of (\ref{eq:DC_vout}), and then use a bisection method to find $\overline{v}_{out}$ satisfying (\ref{eq:DC_vout}). Such $\overline{v}_{out}$ is unique, since the expression on the LHS of  (\ref{eq:DC_vout}) is strictly increasing over $\overline{v}_{out}$.

It is worth noting that the existing   approach to handle the integration on the RHS of (\ref{eq:DC_vout}) is to approximate the inner exponential function with its truncated  Taylor series to the $k_0$-th order as  \vspace{-3mm}
\begin{align} \label{eq:Approx-Exp}
\exp(\dfrac{\sqrt{R_s} y(t)}{\eta V_0})\approx \sum_{k=0}^{k_0}  \dfrac{c_k}{k !} \left(y(t)\right)^k,
\end{align}
where $c_k=(\sqrt{R_s}/\eta V_0)^k$. The approximation in (\ref{eq:Approx-Exp}) is valid when $y(t)$ is  small. Specifically,  $k_0=2$,  known as the {\it linear} model, is commonly adopted in the WPT and  SWIPT/WPC literature (e.g., \cite{515,478}). 
The higher-order Taylor approximation model of the diode with $k_0=4,6$ has been recently considered  in \cite{Clerckx:2016b}.  
In contrast to the above studies, in this paper we avoid the Taylor approximation to  ensure the best accuracy of the nonlinear EH model. 

Last, note that to keep  the output DC voltage $\overline{v}_{out}$ constant over time, the capacitance of the LPF  should be set such that $C R_L \gg T$, explained as follows.  
In the electric circuit of rectenna shown in Fig. \ref{fig:antennafull}, when the diode is reversely biased,  
the output voltage is governed by the discharging law of the capacitor of the LPF, and is proportional to  $\exp(-\tfrac{t}{C R_L})$. 
In this case, by setting e.g.  $C=50 T/R_L$,  the normalized output voltage fluctuation (i.e., divided by the peak of the output voltage) is obtained as $1-\exp(-0.02)=1.98 \%$, which is reasonably small as practically required. 
\section{Problem Formulation and Proposed  Solution}
In this section, we formulate the waveform optimization problem. We then present two approximate solutions to it.
\vspace{-1.5mm}
\subsection{Problem Formulation}
With the result in (\ref{eq:DC_vout}), we now proceed to optimize the sinewave amplitudes and phases, $s_{n}$'s and $\phi_{n}$'s, such that  $\overline{v}_{out}$ is maximized, under the maximum transmit sum-power constraint. The  problem is formulated as \vspace{-1mm}
\begin{align} 
\mathrm{(P0)}:  
\mathop{\mathtt{max}}_{\{s_{n}\ge 0\},\{0\le \phi_{n}<2\pi\}}~& \dfrac{1}{T} \int_{T}\exp(\dfrac{\sqrt{R_s} y(t)}{\eta V_0}) dt \label{eq:P0_Obj} \\
\mathtt{s.t.}~& \sum_{n=1}^{N} s_{n}^2 \le P_T, \label{eq:P0_C1}
\end{align}
with $y(t)$ given in (\ref{eq:y(t)}). 
To ensure that all $\cos(\cdot)$ terms in $y(t)$ are being added constructively to each other, we need to set $\phi_{n}=-\psi_{n}$, $n=1,\ldots,N$. With this optimal phase design, in the rest of this paper, we focus on optimizing the sinewave amplitudes $s_n$'s  by considering  the following problem.\vspace{-1mm}
\begin{align} 
\hspace{-3mm}\mathrm{(P1)}:  
\mathop{\mathtt{max}}_{\{s_{n}\ge 0\}}~&  \dfrac{1}{T} \int_{T} \exp(\dfrac{\sqrt{2R_{s}}  \sum_{n=1}^{N} s_{n} h_{n} \cos (w_n t)}{\eta V_0}) dt\hspace{-1mm} \label{eq:P1_Obj} \\
\mathtt{s.t.}~&  \sum_{n=1}^{N} s_{n}^2 \le P_T. \label{eq:P1_C1}
\end{align}

The objective function and  constraint in  (\ref{eq:P1_Obj}) and (\ref{eq:P1_C1}) are both convex over $s_{n}$'s. 
However, since maximizing a convex function over a convex set is non-convex in general, (P1) is a non-convex optimization problem. 
In the next subsections, we propose two approximate solutions to (P1). 

Note that with the linear model of the diode  or equivalently its second-order truncated Taylor approximation \cite{888}, the waveform optimization problem in (P1) can be simplified as a convex problem and then  solved, where the obtained solution simply allocates all  transmit power to the  frequency tone with the largest channel magnitude, i.e., only  $f_{\hat{n}}$, with $\hat{n}=\arg \max_{n}  h_n$, is being used for WPT.  
While with the higher-order truncated Taylor approximation model of the diode  \cite{Clerckx:2016b}, (P1) can be simplified, but the resulted problem is still non-convex.  In \cite{Clerckx:2016b}, the SCP technique is used to solve such non-convex problem approximately in an iterative manner, where a GP problem needs to be solved at each iteration. 
\vspace{-1mm}
\subsection{Maximal Ratio Transmission (MRT) in  Frequency Tones} \label{sec:MRC_Sol}
Let $z(t)=\exp(\sqrt{2R_{s}}  \sum_{n=1}^{N} s_{n} h_{n} \cos (w_n t)/(\eta V_0))$. 
One approximate solution to (P1) is obtained by replacing the integral in (\ref{eq:P1_Obj}) by the peak value of its integrand $z(t)$  over one period $0\le t \le T$, which is given by   $z(0)=\exp(\sqrt{2 R_{s}} \sum_{n=1}^{N} s_{n} h_{n}  /(\eta V_0))$. 
To justify this approach, we present a numerical example as follows. 
We consider a SISO WPT with the center frequency  $f_c=20$kHz, the total bandwidth $B=2$kHz, and the transmit power limit $P_T=1$W. We set $N=4$, $\Delta_f=0.5$kHz, and $f_0=19$kHz.  
By considering a frequency-flat  channel with $h_n=7 \times 10^{-3}$, $n=1,\ldots, 4$, we equally divide the transmit power over all frequency tones $f_n$'s, i.e., $s_n=1/2$, $n=1,\ldots, 4$. The details about rectenna circuit parameters are given later in Section \ref{Sec:Simul}.  
\begin{figure}
	\centering
	\includegraphics[width=1\linewidth]{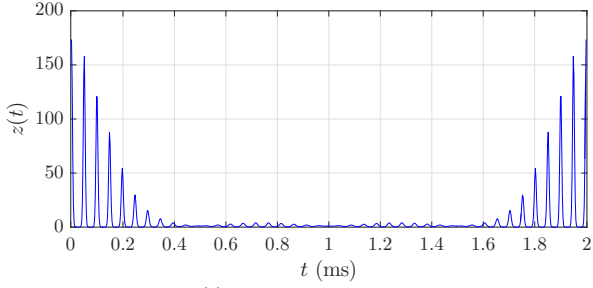} \vspace{-8mm}
	\caption{Illustration of $z(t)$.} \vspace{-3mm}
	\label{fig:Function_z(t)}
\end{figure} 
Accordingly, we plot $z(t)$ in Fig. \ref{fig:Function_z(t)}, from which it is observed that this signal has multiple  large peaks in the vicinity of $t=0$ (and $t=T=2$ms due to the periodicity), but has negligible amplitudes  for rest of time $t$. 
In this case, it seems reasonable to maximize the peak value of $z(t)$ so as to approximately maximize its time average  in (\ref{eq:P1_Obj}). Thus, we reformulate (P1) as follows. \vspace{-1mm}
\begin{align} 
\mathrm{(P2)}:  
\mathop{\mathtt{max}}_{\{s_{n}\ge 0\}}~&   \exp(\dfrac{ \sqrt{2 R_{s}}  \sum_{n=1}^{N} s_{n} h_{n}}{\eta V_0}) \label{eq:P2_Obj} \\
\mathtt{s.t.}~& \sum_{n=1}^{N} s_{n}^2 \le P_T. \label{eq:P2_C1}
\end{align}
(P2) is non-convex, but its optimal solution can be obtained by investigating its KKT conditions.  We omit the details due to space limitation and present its optimal  solution as follows. 
\begin{proposition} \label{Propostion:1}
The optimal solution to (P2) is given by $s_n=h_n \sqrt{P_T/\sum_{k=1}^{N} h_k^2}$, $n=1,\ldots,N$.
\end{proposition}

The sinewave amplitudes $s_n$'s given in Proposition \ref{Propostion:1} can be considered as an approximate  solution to the original problem (P1). Moreover, this solution  is analogous to the MRT based beamforming in  wireless communication, but applied in frequency domain instead of spatial domain. 
\vspace{-2mm}
\subsection{Sequential Convex Programming (SCP) Based Solution} \label{sec:SCP_Sol}
SCP is an iterative method to solve non-convex problems suboptimally, by leveraging convex optimization techniques. 
Specifically, at each iteration $m$, $m=1,2,\ldots$, we approximate  the objective function in (P1) by a linear function using  its first-order Taylor series to form a convex {\it approximate} optimization problem.
Next, we set the values of decision variables $s_{n}$'s for iteration $m+1$ as the optimal solution to the approximate problem at iteration $m$. 
The algorithm continues until a given stopping criterion is satisfied. 
In the following, we provide details of our SCP based algorithm for (P1).

Let $s_{n}^{(m)}$, $n=1,\ldots,N$, denote the  values of decision variables at the beginning of iteration $m$. 
We approximate the objective function in (\ref{eq:P1_Obj}) via its first-order Taylor series as \vspace{-1mm}
\begin{align} \label{eq:Approx_Obj_P1}
\beta_0^{(m)} + \sum_{n=1}^{N}  \beta_{n}^{(m)} \left(s_{n}-s_{n}^{(m)} \right),
\end{align}
with the coefficients given by \vspace{-1mm}
\begin{align}
\hspace{-2mm}&\beta_0^{(m)}\hspace{-1mm}=\hspace{-1mm}\dfrac{1}{T} \hspace{-1mm} \int_{T} \hspace{-1mm} z^{(m)}(t)  dt, \label{eq:beta0}\\ 
\hspace{-2mm}&\beta_{n}^{(m)}\hspace{-1mm}=\hspace{-1mm}\dfrac{1}{T} \hspace{-1mm} \int_{T} \hspace{-1mm} \dfrac{\sqrt{2R_s}}{\eta V_0} h_{n} \cos(w_n t) z^{(m)}(t) dt, \hspace{-0.2mm} ~n\hspace{-1mm}=\hspace{-1mm}1,\ldots,N, \hspace{-1mm}\label{eq:betan}
\end{align}
where  $z^{(m)}(t) \triangleq  \exp(\sqrt{2R_{s}}  \sum_{n=1}^{N} s_{n}^{(m)} h_{n} \cos (w_n t)/ (\eta V_0))$. 
Since  the objective function in (\ref{eq:P1_Obj}) is convex over $s_n$'s, the linear approximation given in (\ref{eq:Approx-Exp}) is  its \textit{global} under-estimator \cite{202}. 
The integrals in (\ref{eq:beta0}) and (\ref{eq:betan}) can be computed numerically  as follows.  
Let $Q\ge 1$ be a large positive integer representing the number of  sub-intervals of equal width used for signal sampling  within the  period $T$, where 
$\Delta_t= T/Q$ denotes the duration of each sub-interval. 
By using the $2$-point closed Newton-Cotes formula  (also known as the trapezoidal rule) \cite{Ube:1997}  to compute the integrals over each sub-interval $(q-1)\Delta_t \le t \le q \Delta_t$, $q\in \{1,\ldots, Q\}$, and then adding up all the results, we can approximate (\ref{eq:beta0}) and (\ref{eq:betan}) as \vspace{-2mm}
\begin{align}
&\tilde{\beta}_0^{(m)}=\dfrac{1}{Q}\sum_{q=1}^{Q} z^{(m)}(q\Delta_t), \label{eq:beta0-sampled}\\ 
&\tilde{\beta}_{n}^{(m)}=\dfrac{1}{Q}\sum_{q=1}^{Q} \dfrac{\sqrt{2R_s}}{\eta V_0} h_{n} \cos(w_n q \Delta_t) z^{(m)}(q\Delta_t), \label{eq:betan-sampled}
\end{align} 
for $n=1,\ldots,N$. Note that in the above,  we have sampled each $z^{(m)} (t)$ at $Q$ equally  spaced points.
Let  $E_0^{(m)}=|\beta_0^{(m)} -\tilde{\beta}_0^{(m)}|$ and $E_n^{(m)}=|\beta_0^{(m)} -\tilde{\beta}_0^{(m)}|$, $n=1,\ldots,N$, denote the approximation errors.  
It can be shown that the error terms, $E_0^{(m)}$ and  $E_n^{(m)}$'s, are all upper-bounded \cite{Ube:1997}. 
Specifically, we have $|E_0^{(m)}|\le \hat{E}_0^{(m)}$ and $|E_n^{(m)}|\le \hat{E}_n^{(m)}$, $n=1,\ldots,N$,  with  $\hat{E}_0^{(m)}=\tfrac{1}{12} (\tfrac{T}{Q})^2 \max_{0 \le t \le T} |\partial^2 z^{(m)}(t)/\partial t^2|$ and $\hat{E}_n^{(m)}=\tfrac{1}{12} \hspace{0.5mm} (\tfrac{h_n \sqrt{2R_s}}{\eta V_0}) \hspace{0.5mm} (\tfrac{T}{Q})^2 \hspace{0.3mm} \max_{0 \le t \le T} |\partial^2 \cos(w_n t) z^{(m)}(t) /\partial t^2|$. 
It is observed that the errors  are quadratically  decreasing over $Q$.  Hence, by setting $Q$ sufficiently large,  high-accuracy approximation  can be achieved.  
In our algorithm, we set $Q=20f_c$ to obtain  $\tilde{\beta}_0^{(m)}$ and $\tilde{\beta}_s^{(m)}$'s.\footnote{To evaluate  the actual value of the objective function of (P1) in (\ref{eq:P1_Obj}) with our obtained solutions, we use the similar Newton-Cotes formula  \cite{Ube:1997}, but with a much larger number  $Q=100f_c$ of samples per period to achieve the best accuracy.}  

Next, we present the approximate problem of (P1) for each iteration $m$ as follows.\vspace{-3mm}
\begin{align} 
\mathrm{(P1-\mathnormal{m})}:  
\mathop{\mathtt{max}}_{\{s_{n}\ge 0\}}~&\tilde{\beta}_0^{(m)} + \sum_{n=1}^{N} \tilde{\beta}_{n}^{(m)} \left(s_{n}-s_{n}^{(m)} \right) \label{eq:P1-l_Obj} \\
\mathtt{s.t.}~& \sum_{n=1}^{N} s_{n}^2 \le P_T. \label{eq:P1-l_C1} 
\end{align}
(P1$-\mathnormal{m}$) is a convex  quadratically constrained linear programming (QCLP). The optimal solution to (P1$-\mathnormal{m}$) is given in the following proposition in closed-form. 
\begin{proposition} \label{Proposition:2}
The optimal solution to (P1$-\mathnormal{m}$) is given by $s_n= \tilde{\beta}_{n}^{(m)} \sqrt{P_T/\sum_{k=1}^{N} (\tilde{\beta}_{k}^{(m)})^2}$, $n=1,\ldots, N$.
\end{proposition}

We set $s_{n}^{(m+1)}$'s according to Proposition \ref{Proposition:2}.  
We can then compute $\tilde{\beta}_0^{(m+1)}$, and update   $\Delta_{\tilde{\beta}_0}=|\tilde{\beta}_0^{(m+1)}- \tilde{\beta}_0^{(m)}|/\tilde{\beta}_0^{(m)}$. 
Let $\epsilon>0$ denote a presumed stopping threshold. 
If $\Delta_{\tilde{\beta}_0} \le \epsilon$, then the algorithm terminates. Otherwise, if $\Delta_{\tilde{\beta}_0} >\epsilon$, then the algorithm will continue to the next iteration.
 
The above iterative algorithm is summarized in Table \ref{Tabel:SCP}, named SCP-QCLP algorithm. This algorithm cannot guarantee to converge to the optimal solution of the waveform optimization problem (P1), but
can yield a point fulfilling the KKT conditions of (P1) \cite{202}.  Hence,  SCP-QCLP algorithm  returns (at least) a locally optimal solution to (P1).
\begin{table}[t!]
	\begin{center} 
		\caption{Proposed iterative algorithm for (P1).} \vspace{-.5mm} \scriptsize{\hrule \vspace{0.05cm} SCP-QCLP Algorithm  \vspace{0.05cm}\hrule  
			\begin{itemize}
				\item[a)]  Initialize $m=1$,    $\epsilon>0$, $\Delta_{\tilde{\beta}_0}>\epsilon$, $Q>0$, and $s_{n}^{(1)}=\sqrt{P_T/N}$, $n=1,\ldots,N$, i.e., equal power allocation over all frequency tones. 
				\item[b)]  {\bf While} $\Delta_{\tilde{\beta}_0}>\epsilon$ \textbf{do}: 
				\begin{itemize}
					\item[$\bullet$] Compute the coefficients $\tilde{\beta}_0^{(m)}$ and $\tilde{\beta}_{n}^{(m)}$, $n=1,\ldots,N$, in  (\ref{eq:beta0-sampled}) and (\ref{eq:betan-sampled}), respectively.  
					\item[$\bullet$] Find the optimal solution to (P1$-\mathnormal{m}$) using  Proposition \ref{Proposition:2}, and set it as $s_{n}^{(m+1)}$. 
					\item[$\bullet$] Update  $\Delta_{\tilde{\beta}_0}=|\tilde{\beta}_0^{(m+1)}- \tilde{\beta}_0^{(m)}|/\tilde{\beta}_0^{(m)}$. 
					\item[$\bullet$] Set $m=m+1$. 
				\end{itemize}
				\item[d)] Return  $s_{n}^{(m)}$'s  as the solution to (P1).
			\end{itemize}
			\hrule  \label{Tabel:SCP}}
	\end{center} \vspace{-6mm}
\end{table}
\vspace{-1mm}
\section{Simulation Results} \label{Sec:Simul}
\begin{figure}
	\centering
	\includegraphics[width=1\linewidth]{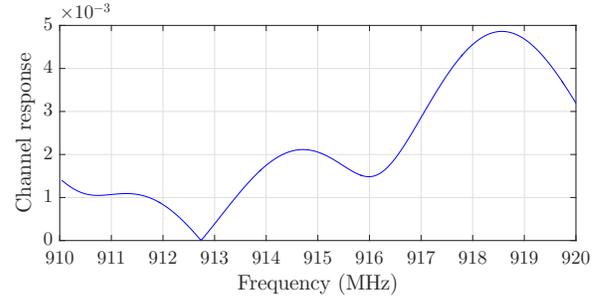} \vspace{-8mm}
	\caption{Channel frequency response.} 
	\label{fig:channelResponse} \vspace{-3mm}
\end{figure}

Consider a SISO WPT system, with central frequency $f_c=915$MHz and  total bandwidth  $B=10$MHz. 
For the channel from the energy transmitter to energy receiver, we  assume $51.67$dB path loss (i.e., they are separated by $10$ meters) in a large open space environment and a NLoS channel power delay profile with $L=18$ paths. 
For simplicity, we assume that the signal power is equally divided among all different paths.  
However, the delay of each path and its phase are assumed to be  uniformly distributed over $[0,~0.3]$ in $\mu$s and $[0,~2\pi]$, respectively.  
Fig. \ref{fig:channelResponse} shows one realization of the frequency response of the assumed channel, which will be used in the following simulations.  
\begin{figure}[t!] \vspace{-1mm}
	\centering
	\includegraphics[width=1\linewidth]{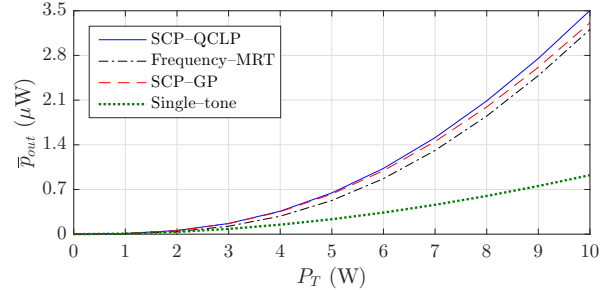} \vspace{-8mm}
	\caption{Deliverable DC power versus the maximum transmit power, with $N=16$.}
	\label{fig:Vdc_vs_PT} \vspace{-3mm}
\end{figure}
\begin{figure}[t!]
	\centering
	\includegraphics[width=1\linewidth]{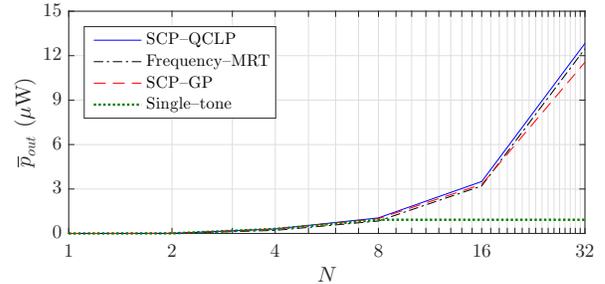} \vspace{-8mm}
	\caption{Deliverable DC power  versus the number of frequency tones used for WPT,  with $P_T=10$W.}   \vspace{-1mm}
	\label{fig:Vdc_vs_N}
\end{figure}
\begin{table}[t!]
	\centering
	\caption{Convergence time comparison.} \vspace{-3mm}
	\label{tab:perf_compar}
	\begin{tabular}{|c|S|S|S|}
		\hline
		\multirow{2}{*}{Design  approach} & \multicolumn{3}{c|}{Convergence time (second)} \\ \cline{2-4} 
		& $N=2$ & $N=8$ & $N=32$ \\ \hline
		SCP-QCLP & 0.014 & 0.033 & 0.425 \\ \hline
		SCP-GP~~~~~& 12.75 & 23.97 & 129.83\\ \hline
	\end{tabular} \vspace{-2mm}
\end{table}
For the rectenna, the ohmic resistance of its antenna is  set as $R_s=50\Omega$, the parameters of its rectifier are given by $I_0=5\mu$A, $V_0=25.86$mV, and $\eta=1.05$, and its load resistance is set as $R_L=1.6$k$\Omega$, same as in \cite{Clerckx:2016b}.  
For SCP-QCLP algorithm in Table I, we set  $\epsilon=10^{-3}$.

For comparison with our proposed frequency-MRT  solution and  SCP-QCLP algorithm,  we consider two other benchmark designs: i) single-tone power allocation based on the linear model of the diode  \cite{888}; and ii) the SCP-GP algorithm \cite{Clerckx:2016b} based on the higher-order  truncated Taylor approximation model of the diode.  
To implement SCP-GP algorithm,  we consider the $4$-th order Taylor approximation model of the diode and set the relative error threshold  as $\epsilon=10^{-3}$, the same as that considered for our SCP-QCLP algorithm. 
Moreover,  we set the same initial point for both algorithms as $s_{n}^{(1)}=\sqrt{P_T/N}$, $n=1,\ldots,N$. 

First, we fix $N=16$.  
By varying the maximum transmit power limit $P_T$,  we plot the  deliverable  DC power $\overline{p}_{out}$  in (\ref{eq:pdc}) under different waveform design schemes in Fig. \ref{fig:Vdc_vs_PT}.  
It is observed that SCP-QCLP achieves the best performance over all values of  $P_T$.   
It is also observed that the frequency-MRT  solution considerably outperforms  the conventional single-tone design for liner EH model. 
It is further observed that the gap between SCP-QCLP and SCP-GP  increases with $P_T$,  explained as follows. 
In \cite{Clerckx:2016b}, SCP-GP is proposed based on the truncated Taylor approximation model of the diode, which is valid when the voltage drop across the diode, i.e., $v_d(t)$ shown in Fig. \ref{fig:antennafull}, is small. 
However, by increasing the transmit power, the peak voltage collected by the rectenna increases, which causes  the voltage drop across the diode to increase.  
As a result, the truncated Taylor approximation model of the diode becomes less accurate, and the performance  of SCP-GP degrades.  
Furthermore, it is observed that the frequency-MRT solution achieves very close performance to both SCP-QCLP and SCP-GP, thus providing a practically  appealing alternative design considering its low complexity.     

Next, we fix $P_T=10$W. By varying the number of utilized frequency tones $N$ for WPT, we plot the  deliverable  DC power $\overline{p}_{out}$  under different  waveform design schemes in Fig. \ref{fig:Vdc_vs_N}.  The convergence time of SCP-QCLP and SCP-GP is also  compared in Table \ref{tab:perf_compar}.\footnote{Simulations are implemented on MATLAB R2015b and tested on a PC with a Core i7-2600 CPU, 8-GB of RAM, and Windows 10.} 
It is observed that  SCP-QCLP achieves the best performance over all values of $N$, and also converges remarkably faster than SCP-GP. 
This is due to the fact that  SCP-QCLP requires only  to solve a simple QCLP problem (with the  optimal solution shown in closed-form in Proposition \ref{Proposition:2})  at each iteration, while SCP-GP needs to solve a GP problem at each iteration, which requires more computational time.  
It is also observed that when $N$ increases, the frequency-MRT solution outperforms that by SCP-GP.

\begin{figure} [t!]
	\begin{center}
		\subfigure[Sinewave amplitudes]
		{\scalebox{0.325}{\hspace{-10mm}\includegraphics{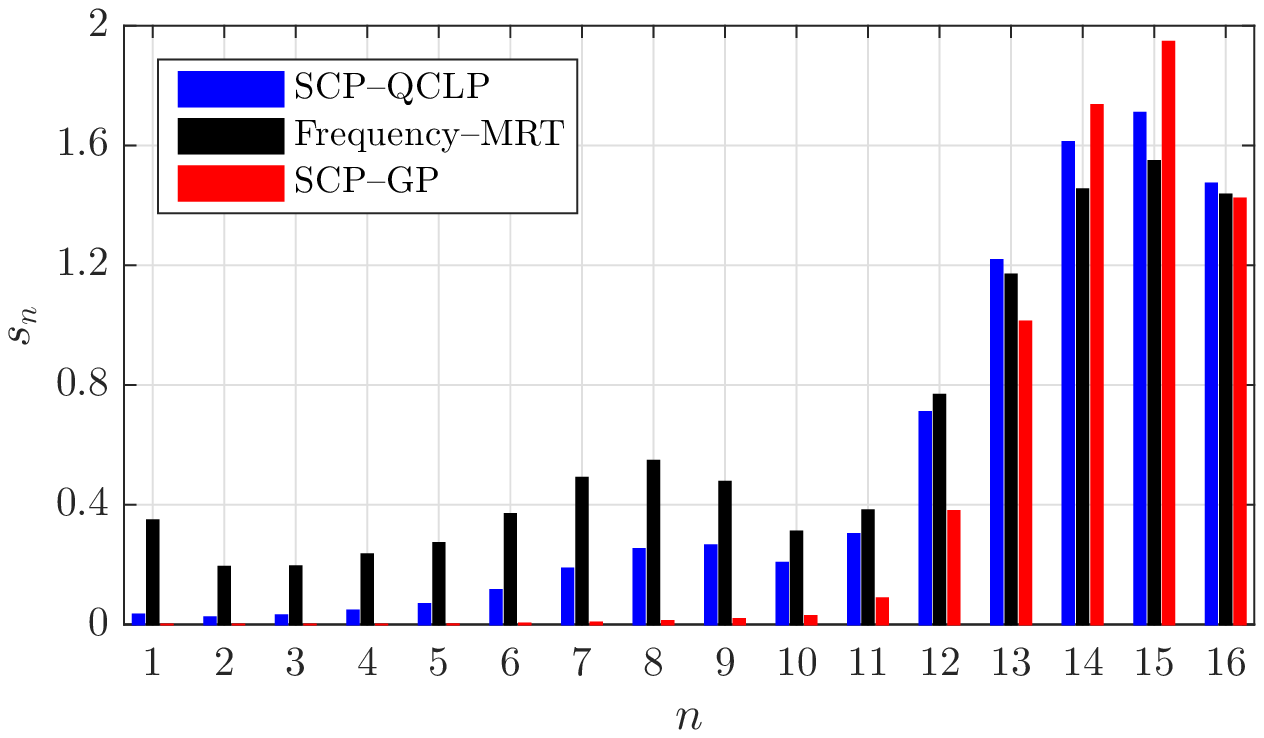}}}  \hspace{-5mm}
		\subfigure[SCP-QCLP]
		{\scalebox{0.325}{\includegraphics{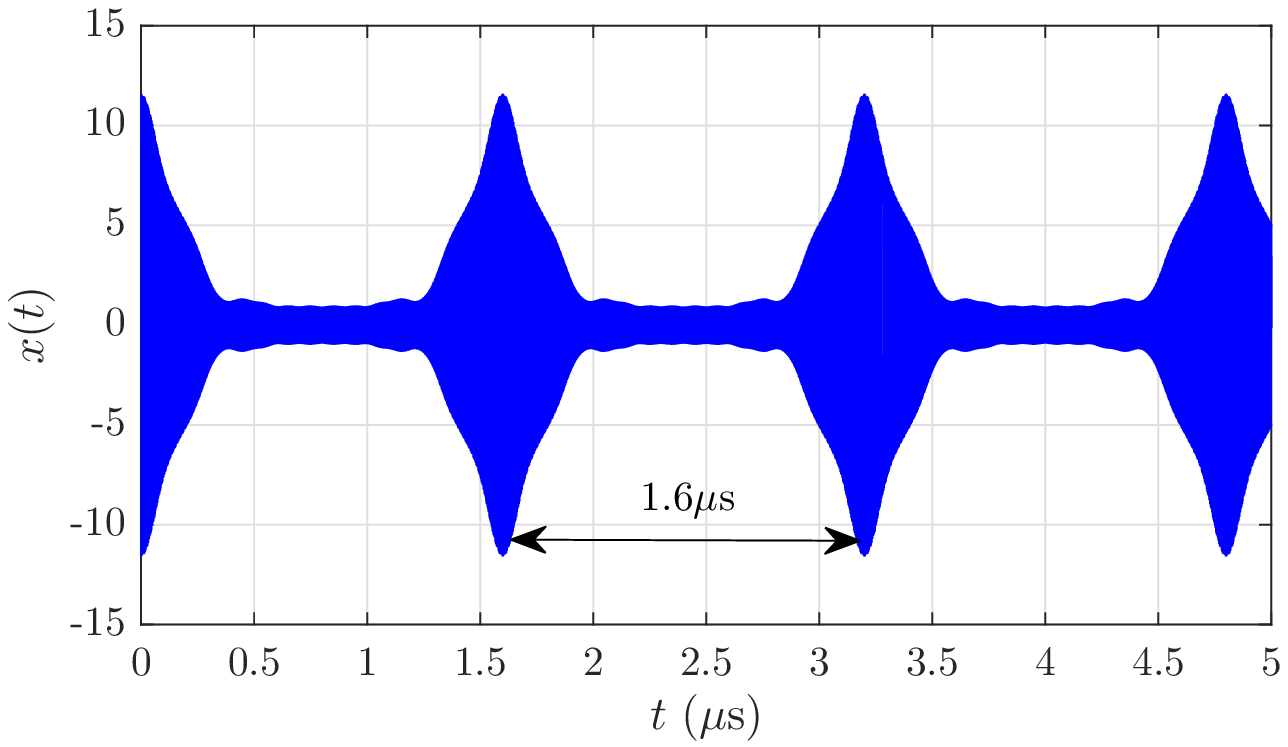}} \hspace{-5mm}}
		\subfigure[Freqeuncy-MRT]
		{\scalebox{0.325}{\hspace{-10mm}\includegraphics{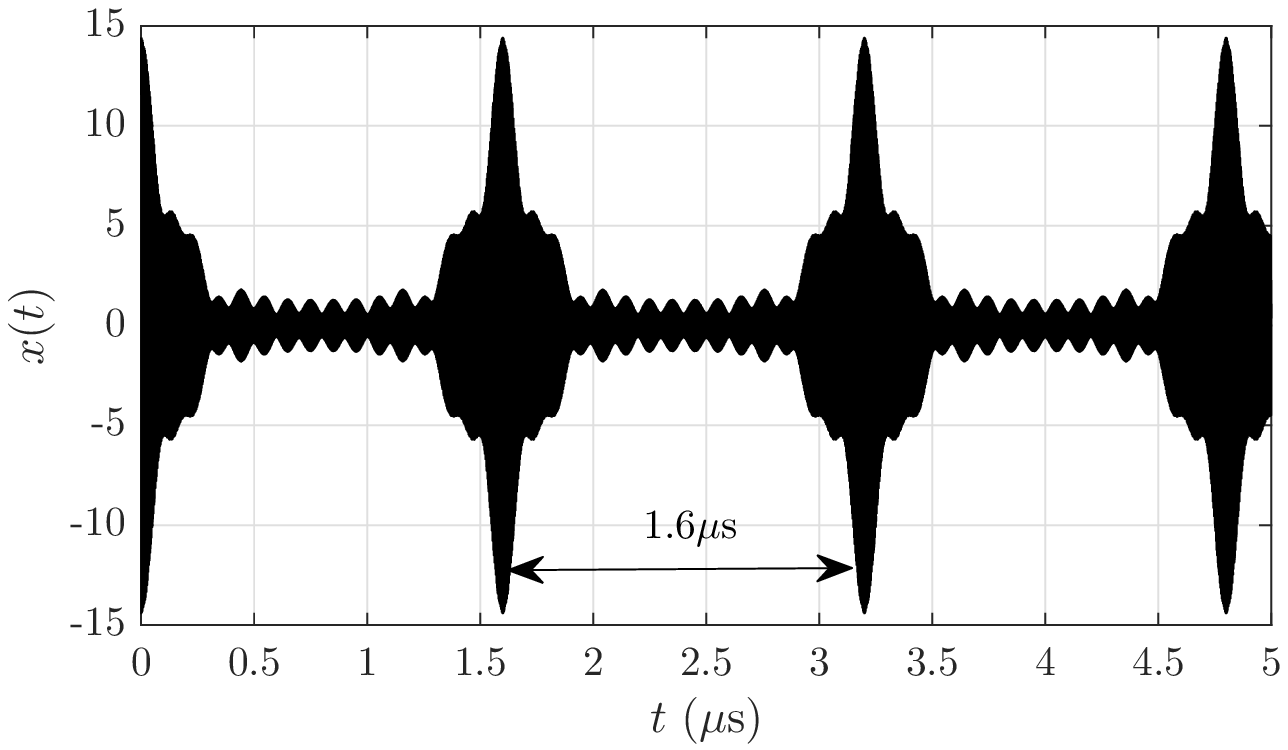}}}\hspace{-5mm}
		\subfigure[SCP-GP]
		{\scalebox{0.325}{\includegraphics{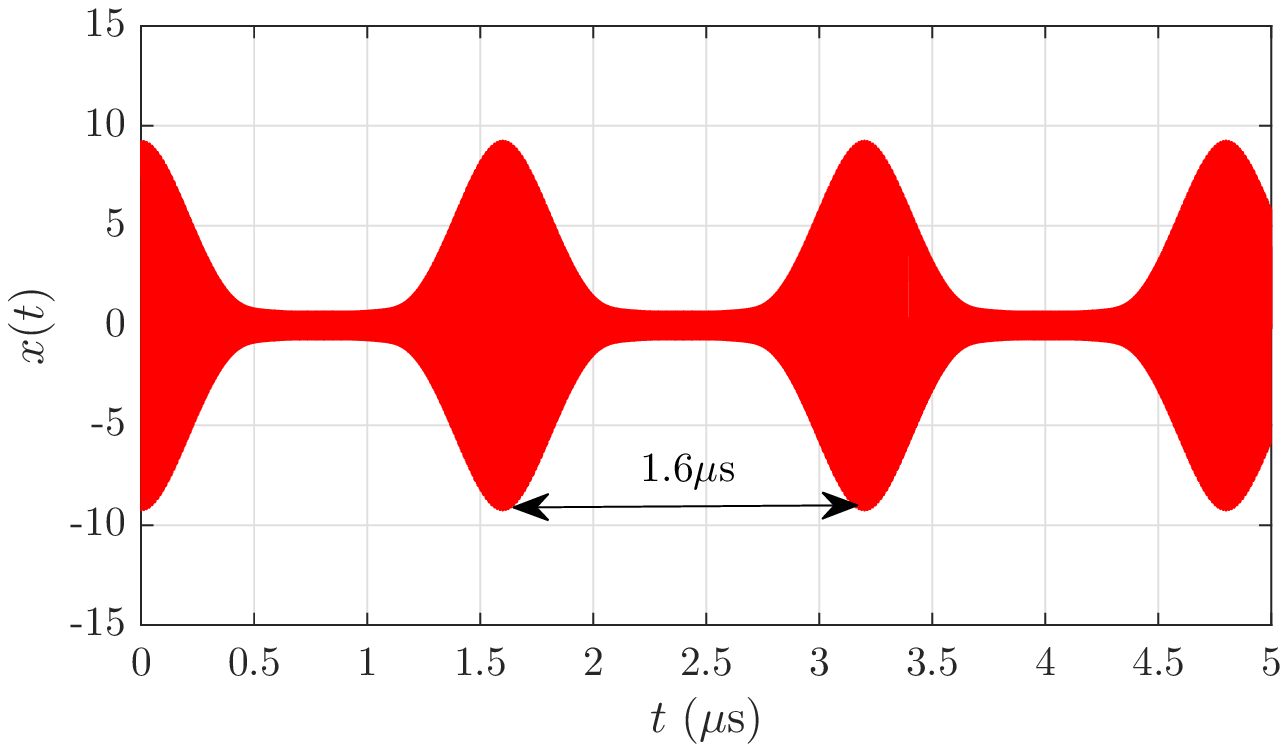}} \hspace{-5mm}}
	\end{center} \vspace{-3mm}
	\caption{Comparison of the sinewave amplitudes obtained via different design schemes as well as their  resulted transmit signal waveform.}
	\label{fig:sn-x(t)} \vspace{-4mm}
\end{figure}

Last, with $P_T=10$W and $N=16$ fixed, we plot the optimized sinewave amplitudes $s_n$'s for different designs as well as their corresponding  signal waveform $x(t)$ in Figs. \ref{fig:sn-x(t)}(a)--(d), respectively. 
It is observed that $x(t)$ under the frequency-MRT solution achieves the largest PAPR compared to the other two designs; however it has been shown in Figs. \ref{fig:Vdc_vs_PT} and \ref{fig:Vdc_vs_N} that its deliverable DC power is less than those by  SCP-QCLP and SCP-GP (when $N$ is not large). 
This is due to the fact that the deliverable DC voltage/power is proportional to the time average of the exponential function of the received RF signal (see (\ref{eq:DC_vout})), but not depending on its peak value solely.
\vspace{-1mm}
\section{Conclusion}
In this paper, we studied the waveform design problem for a SISO WPT system under frequency-selective channels assuming perfect CSI at energy transmitter. 
We  developed a generic EH model based on circuit analysis that accurately captures the rectenna nonlinearity without relying on its Taylor approximation. 
Based on this model, we formulated a new multisine energy waveform optimization problem subject to a given transmit power constraint. Although the formulated problem is non-convex, we proposed two suboptimal solutions for it with low complexity. Simulation results showed the superiority of our proposed waveform solutions over the existing designs based on the truncated Taylor approximation in terms of both performance and computational time. 
\vspace{-1mm}
\bibliographystyle{IEEEtran} 
\bibliography{IEEEabrv,IEEEfull}

\end{document}